\documentclass[preprint,aps,nofootinbib]{revtex4}
\usepackage{graphicx}
\usepackage{epstopdf}
\usepackage{subfigure}
\usepackage{textcomp}
\usepackage{dsfont}
\usepackage{amsfonts}
\usepackage{amsmath}
\usepackage{relsize}
\usepackage{amssymb}
\begin{document}

\title{Neutron stars, ungravity, and the I-Love-Q relations}

\author{Hodjat Mariji\footnote{Present Address: Departamento de Engenharia Electrot\'{e}cnica, Institudo de Telecomunica\c{c}\~{o}es, Universidade de Coimbra, P\'olo II, 3030-290 \: Coimbra, Portugal. \\	Email: astrohodjat@gmail.com, \quad  hodjat.mariji@uc.pt}}
\author{Orfeu Bertolami\footnote{E-mail: orfeu.bertolami@fc.up.pt}}

\vskip 0.3cm

\affiliation{Departamento de F\'\i sica e Astronomia, Faculdade de Ci\^encias da Universidade do Porto and Centro de F\'\i sica do Porto\\
Rua do Campo Alegre 687, 4169-007 Porto, Portugal}

\vskip 0.5cm

\begin{abstract}

\vskip 0.5cm

{
 We study neutron stars (NSs) in an ungravity (UG) inspired model. We examine the UG effects on the static properties of the selected NSs, in different mass and radius regimes, i.e., ultralow, moderate, and ultrahigh NSs, using a polytropic equation of state approach. Based on the observational data, we obtain bounds on the characteristic length and scaling dimension of the UG model. Furthermore, we obtain dynamic properties, such as inertial moment (I), Love number (Love), and quadrupole moment (Q) of a slowly rotating NS in the presence of the exterior gravity and ungravity fields. The UG model is also examined with respect to the I-Love-Q universal relation.
}

\vskip 0.5cm


\vskip 0.3cm

\textbf{PACS Number(s):} 04.20.Fy, 04.80.Cc, 04.25.Nx

\end{abstract}

\maketitle

\section{Introduction}

Recently, there has been a renewed interest in the study of neutron stars (NSs). Being the densest observable objects in the Universe, NSs are good cosmic laboratories for astrophysicists, for instance, to predict conditions for the formation of black holes \cite{1}, to investigate gravitational waves, especially through the study of NS mergers \cite{2}, and to test gravity models \cite{3, 4}. Although the equation of state (EoS) and the composition from core to crust of NSs play an important role in the study of their properties  \cite{5, 6, 7, 8}, recent work  on the calculation of the dynamic features, such as the inertial moment (I), the tidal Love number (Love), and the quadrupole moment (Q), shows that they are independent of the EoS and of the composition (see Refs. \cite{9, 10} and refs. therein). By combining the universal I-Love-Q relations with the information from gravitational wave measurements of binary pulsars (advanced LIGO \cite{11}, Virgo \cite{12}, and KAGRA \cite{13}), one can get a unique, model-independent and internal-structure independent test of General Relativity \cite{9, 14} and obtain information on the EoS \cite{15, 16}. Regarding the universality of the I-Love-Q relations, it is also interesting to examine gravity models for which the dynamic features of NSs are altered. One interesting example is $ungravity$ (UG) \cite{17}. UG arises from the assumption of coupling between spin-2 unparticles (UPs) and the stress-energy tensor \cite{18}. The UP idea has been prepared in order to introduce scale invariance at the low-energy sector of the Standard Model \cite{18, 19}. Since the scale dimensions of the UP operators can take non-integral values, this leads to peculiar features in the energy distributions of some processes involving Standard Model particles. Further investigations of UP effects have immediately been carried out in collider physics and elsewhere \cite{20, 21, 22, 23}.
The effect of UP states in astrophysics and cosmology have been extensively studied \cite{24, 25, 26, 27, 28, 29, 30, 31, 32, 33, 34}. Recently, the effect of an UG-inspired model on the properties of the Sun has been considered and astrophysical bounds on the UG parameters, i.e., the scaling dimension and length scale, have been obtained \cite{35}. More recently, using the UG-inspired model and the polytropic and degenerate gas approaches, there have been obtained bounds on the UG parameters, i.e., the scaling dimension and length scale, based on the observational data of white dwarfs, and have been found white dwarfs with masses above the Chandrasekhar limit \cite{36, 37}.

In this work, using the polytropic EoS approach, we study the properties of NSs in the framework of an UG-inspired model \cite {35, 36}. The motivation for this work is the application of the universal I-Love-Q relation to examine the UG model at the Newtonian limit. We also aim to investigate the astronomical constrains on the UG parameters, i.e., the scaling dimension and the characteristic length, with respect to observational data of NSs. Based on the observational data of the selected pulsars, i.e., 4U1746-37 \cite{38}, M13 \cite{39}, and J0348+0432 \cite{40}, at the ultralow, moderate, and ultrahigh mass regimes, respectively, we can get bounds on the UG parameters. We shall analytically show how the UG I-Love-Q relations deviate from the universal I-Love-Q relations for slowly rotating NSs with uniform density. Additionally, we also show that there is a deviation between the UG I-Love-Q and I-Love-Q relations when the polytropic index is unity and the UG scaling dimension is very close to unity. Hence, astrophysical bounds on UG parameters can be obtained according to universal I-Love-Q relations at the Newtonian limit. This paper is organized as follows: in section II, the UG model is concisely explained and the UG version of the Newtonian hydrostatic equilibrium equations are presented; in section III, the UG I-Love-Q relations are presented by considering a slowly rotating NS in the presence of a gravitational field of a partner. Finally, our results are presented and discussed in section IV.

\section{The UG Hydrostatic Equilibrium Equation}

In the UG model framework \cite {17}, a modification of the Newtonian gravitational potential is introduced through the coupling of spin-2 unparticles $O^{U}_{\mu\nu}$ \cite{18, 19} to the stress-energy tensor of Standard Model states, $T^{\mu\nu}$. The resulting stress-energy tensor has the following form \cite {17}:
\begin{equation}\label{1}
\mathcal{T}^{\mu\nu}=T^{\mu\nu}+\left({\kappa_{*} \over \Lambda^{d_{U}-1}_{U}}\right) g^{\mu\nu}T^{\sigma\rho}O^{U}_{\sigma\rho},
\end{equation}
where $d_{U}$ and $\Lambda_{U}$ ($\geq 1 \: TeV$) are the scaling dimension and the energy scale of $O^{U}$, respectively. It is worth pointing out that the lower bound of $\Lambda_{U}$ refers the lack of detection of these interactions within the available energy range. In Eq. (\ref{1}), $\kappa_{*}=\Lambda^{-1}_{U}\left( \Lambda_{U}\over M_{U}\right)^{d_{UV}}$ where $M_{U}$ is the large mass scale and $d_{UV}$ is the dimension of the hidden sector operators of the ultraviolet theory which possess an infrared fixed point \cite{17}. 
It is also worth mentioning that constraints on the UG parameters are obtained through astrophysical and cosmological arguments. Based on the precision submillimeter tests of the gravitational inverse square law \cite{41}, the UG model has been constrained at a short distance and the allowed regions are obtained for $R_{*}$ as a function of $d_{U}$ and $M_{U}-\Lambda_{U}$ parameter space for various values of $d_{U}$ \cite{17}. Ungravity in the Newtonian limit was shown to have no impact on the claimed fly-by anomaly \cite{42}. Deviations in planetary orbits and perihelion precession were considered in Refs. \cite{29, 43}, Constraints from Big Bang Nucleosynthesis were discussed in Ref. \cite{33}, and implications for dark energy and entropic gravity in Refs. \cite{44, 45}, respectively.

In order to compute the effects of the UPs to the lowest order correction to the Newtonian gravitational potential, the metric $g^{\mu\nu}$ is replaced by the Minkowski metric $\eta^{\mu\nu}$ in Eq. (\ref{1}). The resulting Newtonian gravitational potential in the UG model framework is given by \cite {17}
\begin{align}\label{2}
 \phi_{*}(r) = \phi(r) \phi_{\alpha}(r), && \phi_{\alpha}(r)=G_{\alpha}\left[1+ \left( {R_{*} \over r}\right) ^{\alpha-1}\right]
\end{align}
where $\phi(r)=-{GM\over r}$ is the Newtonian gravitational potential, $R_{*}$ is the length scale which characterizes the UG interactions, and $\alpha$ is associated with $d_{U}$ through $\alpha=2d_{U}-1$. In Eq. (\ref{2}), $G_{\alpha}$, the gravitational constant coefficient of UG, is given by
\begin{equation}\label{3}
 G_{\alpha}= {1\over 1+\left( {R_{*}\over R_{0}}\right)^{\alpha-1}},
\end{equation}
where $ R_{0} $ is the distance in which the UG potential, $\phi_{*}$, matches onto the Newtonian one. It is obvious, from Eqs. (\ref{2}) and (\ref{3}), that by choosing $\alpha=1$, we obtain $\phi_{\alpha}=1$ and then we can recover the ordinary Newtonian gravitational potential. As a good approximation, by considering the value of $\alpha$ near unity, we can write $G_{\alpha}\simeq 1/2$. Without loss of generality we set this approximation which allows for obtaining the bounds on the relevant parameters of the UG model. 
Even though UG models were originally conceived to understand the effects of UPs at very short distances, we shall consider UG models phenomenologically in order to introduce corrections to Newtonian gravity at astrophysically interesting scales.  
  In order to study the effect of UG on NSs, we consider the hydrostatic equilibrium equation for a perfect fluid at the Newtonian limit (NHE). In this case, the most general Tolman-Oppenheimer-Volkoff equation \cite{46},
\begin{equation}\label{4}
 {dP(r)\over dr}=-{GM(r)\rho(r)\over r^{2}}\left[1+{P(r)\over \rho(r)c^{2}}\right]\left[1+{4\pi r^{3}P(r)\over M(r)c^{2}}\right]\left[1-{2GM(r)\over c^{2}r}\right]^{-1},
\end{equation}
gets reduced to the NHE equation as $P(r)\ll \rho(r)c^{2}$, $r^{3} P(r)\ll M(r) c^{2}$, and $GM(r)\ll c^{2}r$:
\begin{equation}\label{5}
 {dP(r)\over dr} = -\rho(r){ d\phi(r) \over dr },
\end{equation}  
where ${GM(r)\over r^{2}}$ has been replaced by ${ d\phi(r) \over dr }$.

In order to obtain UGHE, we consider a mass element, $\delta M_{*}(r)=\rho_{*}(r)dr\delta S$, of the Newtonian static fluid ball within the concentric sphere of radius $r$ and thickness $dr$ in the presence of the UG potential, $\phi_{*}(r)$. The mass element is submitted to the ungravity force, $-[d\phi_{*}(r)/dr]\delta M_{*}(r)$, which is in equilibrium with the pressure force, $-[dP_{*}(r)/dr]dr \delta S$, resulting from the gas pressure difference between $r$ and $r+dr$. Straightforward mathematics leads to UGHE as follows:
\begin{equation}\label{6}
 {dP_{*}(r)\over dr}=-\rho_{*}(r){d\phi_{*}(r)\over dr}
\end{equation}
where the subscript $*$ indicates that the quantities are calculated in the presence of UG. It is clear that setting $\alpha=1$ in the UGHE equation leads to the NHE equation, Eq. (\ref{5}). 

 Equation (\ref{6}) and writing $dM_{*}(r)=4\pi\rho_{*}(r)r^{2}dr$ together with boundary conditions, $M_{*}(0)=0$ and $P_{*}(R_{s})=0$, where $R_{s}$ is the radius of the NS, allow for obtaining the radius and mass, $M_{s}$, for different values of $\alpha$ and $R_{*}$ for a given value of core density, $\rho_{c}$, as input. It should be mentioned that we admit values of the UG parameters for which the calculated $M_{s}$ and $R_{s}$ are within the allowed observational intervals. In Sec. IV, we shall present the numerical results for 4U1746-37 \cite{38}, M13 \cite{39}, and J0348+0432 \cite{40}, for NSs in the ultralow, moderate, and ultrahigh mass regimes, respectively. 

\section{The UG I-Love-Q relations}

  We consider a slowly rotating NS which is slightly deformed due to rotation and/or tidal fields. We assume that the time scale for changes of the relevant quantities is sufficiently long so that the NS can be assumed to be in hydrostatic equilibrium. By applying an external gravitational field, $\varepsilon_{*ij}$, to a NS, the resulting multipole moment of its mass distribution, $Q_{*ij}$, is given by \cite{47}
\begin{equation}\label{7}
 Q_{*,ij}=-\lambda_{*} \varepsilon_{*,ij},
\end{equation}
where $\lambda_{*}$ is the Love number and $\varepsilon_{*,ij}={\partial^{2}\phi_{*} \over \partial x^{i}\partial x^{j}}$. Deformation due to rotation and tidal forces of the NS is the same in the Newtonian limit \cite{14, 48}. We keep this feature in the presence of UG, that is, $\lambda_{*,\: rot}=\lambda_{*,\: tid}$, and consider the rotational deformations in this work rather than solving differential equations as has been done in Ref. \cite{14}. In the Newtonian limit, for a spheroid shaped NS rotating around the $z$ axis, the quadrupole moment tensor reads \cite{49}
\begin{equation}\label{8}
 Q_{*, ij}=diag(-{1 \over 3}Q_{*},-{1 \over 3}Q_{*},{2 \over 3}Q_{*}),
\end{equation}
where 
\begin{equation}\label{9}
 Q_{*}=\int \rho_{*}(r')P_{2}(cos\theta')r'^{2}d^{3}r'.
\end{equation}
The inertial moment is given by \cite{50}
\begin{equation}\label{10}
 I_{*}={2 \over 3} \int \rho_{*}(r') r'^{2} d^{3}r'.
\end{equation}
Since in the Newtonian limit, the quadrupolar contribution of the centrifugal potential reduces to $\Omega_{*}^{2}$, the squared angular velocity of the object around the $z$ axis \cite{48}, the rotational Love number is given by $ \lambda_{*}=-{Q_{*} \over \Omega_{*}^{2}}$.
Recalling the dimensionless $\overline{I}$, $\overline{\lambda}$, and $\overline{Q}$ introduced in Ref. \cite{14}, for the UG case we can define
\begin{align}\label{11}
\underline{I}_{*}={I_{*} \over G^2 M_{s,*}^{3}/c^4}, && \underline{\lambda}_{*}={\lambda_{*} \over G^4 M_{s,*}^{5} /c^{10}}, && \underline{Q}_{*}=-{Q_{*} \over G^2 M^{3}_{s,*} \chi_{*}^{2} / c^4},
\end{align}
for the dimensionless UG inertial moment, Love number ($\lambda_{*,\: rot}=\lambda_{*,\: tid}$), and quadrupole moment, respectively. In Eq. (\ref{11}), $\chi_{*}$, is a dimensionless measure of a NS' angular momentum, is defined by $\chi_{*}={I_{*}\Omega_{*} \over G M_{s,*}^{2} / c}$ \cite{50}. Now, we introduce the UG I-Love-Q relations:
\begin{align}\label{12}
\overline{C}_{*,I\lambda}= {\underline{I}_{*} \over \left(\underline{\lambda}_{*} \right)^{2/5}}, && \overline{C}_{*,IQ}= {\underline{I}_{*} \over \left(\underline{Q}_{*} \right)^{2}}, && \overline{C}_{*,Q\lambda}= {\underline{Q}_{*} \over \left(\underline{\lambda}_{*} \right)^{1/5}},
\end{align}

 In the next subsections we analytically present the UG I-Love-Q relations, i.e., $\overline{C}_{*}$'s, in the Newtonian limit for $n=0$ and $n=1$ polytropes and compare them to the ordinary ones, i.e., $\overline{C}$'s \cite{14}.

\subsection{UG I-Love-Q for n=0 polytrope}

For a slowly rotating NS with an uniform density, vanishing polytropic index, that is, $\rho=cte$, we consider a \textit{Maclaurin spheroid} \cite{47} with uniform angular velocity $\Omega_{*}$ and with eccentricity $e=\sqrt{1-{c^{2} \over a^{2}} } $, where $a$ and $c$ are the semimajor and the semiminor axis, respectively. In the framework of the UG model [cf. Eq. (\ref{2})], we assume the gravitational potential at any point inside the \textit{Maclaurin spheroid} NS as follows:
\begin{equation}\label{13}
 \phi_{*,in}(x,y,z)=-\pi \rho_{*} G G_{0\alpha}f^{2}(x,y,z) \left[1+\left({f^{2}(x,y,z)\over R_{*}^{2}}\right)^{\alpha-1}\right],
\end{equation}
where $\rho_{*}=cte$ and $f(x,y,z)=\sqrt{A-A_{1}x^{2}-A_{2}y^{2}-A_{3}z^{2}}$. The constant coefficients $A, A_{1}, A_{2}$, and $A_{3}$ are given by \cite{47}
\begin{align}\label{14}
 A={2a^{2}\sqrt{1-e^{2}} \over e} sin^{-1}e, && A_{1}=A_{2}={A \over 2(ae)^{2}}-{1-e^{2} \over e^{2}}, && A_{3}={2 \over e^{2}}-{A \over (ae)^{2}}.
\end{align}
In Eq.(\ref{13}), we introduce $G_{0\alpha}$ as
\begin{equation}\label{15}
 G_{0 \alpha}={1 \over 1 + {\left( f_{0}^{2} \over R_{*}^{2} \right)^{\alpha-1}}},
\end{equation}
where the constant factor $f_{0}$ is set such that the UG I-Love-Q relations match onto the ordinary ones.

Now, by assuming that the slowly rotating fluid satisfies UGHE, then $ {d\overrightarrow{v_{*}} \over dt}=-{\triangledown P_{*} \over \rho_{*}}-\triangledown \phi_{*}$, where the velocity of the fluid is given by $\overrightarrow{v_{*}}=\overrightarrow{\Omega_{*}}\times \overrightarrow{r}$ and the pressure is introduced by $P_{*}=P_{*,c}\left[ 1- {\left( x^{2}+y^{2}\right) \over a^{2}}-{z^{2} \over c^{2}} \right]$. Setting $\Omega_{*}$ along the $z$ axis and considering the $x$ and $z$  components of ${d\overrightarrow{v_{*}} \over dt}$, we obtain \cite{47}
\begin{align}\label{16}
P_{*,c}= \pi G G_{0\alpha} \rho_{*} c^{2} A_{3}  , && \Omega_{*}^{2} = 2\pi G G_{0\alpha} \rho_{*} \left( A_{1}- A_{3} {c^{2} \over a^{2}} \right).
\end{align}
By expanding $\Omega_{*}$, Eq. (\ref{16}) to $O(e^{3})$, we obtain
\begin{equation}\label{17}
\Omega_{*}=\sqrt{{8\pi GG_{0\alpha}\rho_{*} \over 15 }} e
\end{equation} 
In Eq. (\ref{9}), by considering $0 \leq r'\leq R_{s}(\theta')$, where $R_{s}(\theta')=\left[ {sin^{2}\theta' \over a^{2}}+{cos^{2}\theta' \over c^{2}} \right]^{-1/2} $, and substituting $e$ from Eq. (\ref{17}), we obtain for an incompressible NS in the UG framework
\begin{align}\label{18}
Q_{*}=-{R_{s}^{5}\Omega_{*}^{2} \over 2GG_{0\alpha} }, && \lambda_{*}=-{R_{s}^{5} \over 2GG_{0\alpha} },
\end{align}
for the rotational quadrupole moment and the Love number ($\lambda_{*,\: rot}=\lambda_{*,\: tid}$), respectively.
Since our incompressible NS has a constant density, $I_{*}$ has a same form as the usual case:
\begin{equation}\label{19}
I_{*}= {2 \over 5} M_{s}R_{s}^{2}.
\end{equation}  
Recalling the compactness parameter as $C={GM_{*s} \over R_{*s}c^{2}}$ \cite{14} and using Eqs. (\ref{11}), (\ref{17}), and (\ref{18}), we obtain
\begin{align}\label{20}
\underline{I}_{*}= {2 \over 5}{1 \over C^{2}} , && \underline{\lambda}_{*}= {1 \over G_{0\alpha}}\left( {1 \over 2}{1 \over C^{5}}\right) , && \underline{Q}_{*}= {1 \over G_{0\alpha} } \left( {25 \over 8 }{1 \over C }\right) .
\end{align}
We can see the tidal and rotational effects on a slowing rotating NS in the presence of UG differ from the usual case by the factor $G_{0\alpha}^{-1}$. Thus, from Eqs. (\ref{12}) and (\ref{20}) the UG I-Love-Q relations are given by 
\begin{align}\label{21}
\overline{C}_{*,I\lambda}= {2^{7/5} \over 5} G_{0\alpha}^{2/5}, && \overline{C}_{*,IQ}= {128 \over 3125}G_{0\alpha}^{2} , && \overline{C}_{*,Q\lambda}= {25 \over 2^{14/5}} {1 \over G_{0\alpha}^{4/5}}.
\end{align}
Comparing the UG I-Love-Q relations with those of the Newtonian case [Eqs. (74), (75), and (76) of Ref. \cite{14}], for $ \alpha=1$, we get the same results provided that $G_{0\alpha}=1$, that is, we set $f_{0}=0$. It means, for an incompressible \textit{Maclaurin} type NS in the UG model the gravitational constant should be like the Newtonian gravitational constant.
In Sec. IV we shall further discuss this issue.

\subsection{UG I-Love Q for n=1 polytrope}

In the case of $n=1$ polytrope of a slowly rotating fluid, we consider the Roche model in which the distribution of the bulk of the mass is unchanged by the rotation and, therefore, the gravitational potential remains $\phi=-{GM \over r}$ ($\phi_{*}={-GM \over r}\phi_{\alpha}$ in the UG framework) at the outer layer \cite{47}. Thus, considering the centrifugal potential as $\phi_{*c}=-{1 \over 2} \Omega_{*}^{2}\left(x^{2}+y^{2}\right)$ and keeping the UGHE situation, we have
\begin{align}\label{22}
{1 \over \rho_{*}} \nabla P_{*}\left(r\right) + \nabla \Psi_{*}\left(r\right)=0, &&  \Psi_{*}\left(r\right)=\phi_{*}\left(r\right)+\phi_{*c}\left(r\right)
\end{align}
where $\rho_{*}$ is the solution of Eq. (\ref{6}) in which $P_{*}=K\rho_{*}^{2}$, where $K$ is a constant. After integrating over both sides of Eq. (\ref{22}), setting the same integral constant as the nonrotating case \cite{47}, i.e., $\Psi_{*}\left( R_{s}\right)=\phi_{*}\left( R_{s}\right) $, then after some manipulation (cf. the Appendix A), the maximum value of $\Omega_{*}$ at the equator of NS can be shown to be
\begin{equation}\label{23}
\Omega_{*}^{2}=G_{\alpha}\left[1+\alpha\left({2\over 2+\alpha}\right)^{{\alpha-1\over \alpha}}\left({R_{*}\over R_{s}}\right)^{\alpha-1}\right] \left({2\over2 + \alpha}\right)^{{3\over\alpha}} \left( {GM_{s} \over R_{s}^{3}}\right) .  
\end{equation}
For $\alpha=1$, we obtain $\Omega_{*}^{2}= \left({2 \over 3}\right)^{3}{GM_{s}\over R_{s}^{3}}$ as expected \cite{47}.
Now, we need an analytic expression for $\rho_{*}$ in order to obtain the UG I-Love-Q relations. Since there is no straightforward form for the quantity from the nonlinear UGHE equation, Eq. (\ref{6}), we solve this equation at the limit of $\alpha\rightarrow 1$. It is useful to get the UG Lane-Emden equation in terms of two dimensionless variables, $\theta$, and $\xi$, to express the density and radial distance with respect to the center of the star values: $\rho=\rho_{c} \theta^{n}$ and $r =\beta \xi$, where $\rho_{c}$ is the density at the center of the NS and $\beta=\left ({(n+1)K \over 4\pi G} \right)^{1/2} $ \cite{35}. By including $P=P_{c}\theta^{n}$ with $P_{c}=K\rho_{c}^{{n+1 \over n}}$ in Eq. (\ref{6}), we obtain for $n=1$ \cite{36}
\begin{equation}\label{24}
\theta''+\left[2\xi^{-1}+\left( \alpha-1\right)A_{\alpha}\xi^{-\alpha} \right]\theta'+\left[G_{\alpha}+A_{\alpha}\xi^{-\alpha+1} \right]\theta=0, 
\end{equation}
where $A_{\alpha}=\alpha G_{\alpha} \xi_{*}^{\alpha-1}$ with $\xi_{*}={R_{*} \over \beta}$. Looking for a solution as $\theta \sim \varphi \xi^{-1} e^{{A_{\alpha} \over 2}\xi^{1-\alpha}}$, our nonlinear second-order differential equation gets reduced to
\begin{equation}\label{25}
\varphi''+ \omega_{\alpha}^{2} \varphi=0,
\end{equation}
where
\begin{equation}\label{26}
\omega_{\alpha}^{2} = G_{\alpha}+A_{\alpha} \xi^{-\alpha+1}+{ \alpha \left( \alpha-1 \right) \over 2 } A_{\alpha} \xi^{-\alpha-1}-\left[ {( \alpha-1 ) \over 2 } A_{\alpha} \xi^{-\alpha} \right]^{2}.
\end{equation}
Since we aim to establish how the UG I-Love-Q relations differ from the ordinary ones, we solve Eq. (\ref{25}) in the $\epsilon\rightarrow 0 $ limit, where $\epsilon=\alpha-1$. So, ignoring all terms in Eq. (\ref{26}) with coefficients $\epsilon$, $\epsilon^{2}$, etc., hence 
\begin{equation}\label{27}
\omega_{\alpha}^{2} \simeq G_{\alpha}+{1 \over \alpha}A_{\alpha}\xi^{1-\alpha}.
\end{equation}
In this limit, we can see that $\omega'_{\alpha}=0$. Thus, the solution of Eq. (\ref{25}) is given by $\varphi=A sin(\omega_{\alpha}\xi)+B cos(\omega_{\alpha}\xi)$. In order to satisfy the boundary conditions, i.e., $\theta(0)=1$, the solution of Eq. (\ref{24}) reads
\begin{equation}\label{28}
\theta ={ sin( \omega_{\alpha} \xi ) \over \xi} e^{- {A_{\alpha} \over 2 }  \left(1- \xi^{1-\alpha} \right) }.
\end{equation}
Hence, the density of a slowing rotating $n=1$ polytrope fluid reads
\begin{equation}\label{29}
\rho_{*}=\rho_{c} { sin( \omega_{\alpha} \xi ) \over \xi} e^{- {A_{\alpha} \over 2 }  \left(1- \xi^{1-\alpha} \right) }.
\end{equation}
In order to have a more convenient expression for the density, we consider that $\alpha \approx 1$ and, in turn, $ e^{- { A_{\alpha} \over 2 } \left( 1- \xi^{1-\alpha } \right) } \approx 1$. Thus, we have
\begin{equation}\label{30}
\rho_{*}(r)={\sqrt{2G_{\alpha}} \over 4} \rho_{c} \left[ 3+ \left( {R_{*} \over r}\right)^{\alpha-1} \right]. 
\end{equation}
In the above equation, we get $\rho_{*}(r)=\rho_{c}$ by setting $\alpha=1$. It means that in the $\alpha \approx 1$ limit, a slowly rotating $n=1$ polytrope fluid is close to the $n=0$ one. 

Now, using Eqs. (\ref{9}), (\ref{10}), and (\ref{30}), we obtain  
\begin{equation}\label{31}
I_{*}={2 \over 5}M_{s}R_{s}^{2} \left[ {1+{5/3 \over 6-\alpha} \left( {R_{*} \over R_{s}}\right)^{\alpha-1} \over 1+ {1 \over 4-\alpha} \left( {R_{*} \over R_{s}}\right)^{\alpha-1}} \right],
\end{equation}
for the inertial moment and 
\begin{equation}\label{32}
Q_{*}=-{4\pi \over 5}{\sqrt{2G_{\alpha}} \over 4} \rho_{c} R_{s}^{5} \left[ 1+{5/3 \over 6-\alpha} \left( {R_{*} \over R_{s}}\right)^{\alpha-1} \right]e^{2}
\end{equation}
for the quadrupole moment, ignoring $O(e^{3})$ terms. Since we have estimated the dynamic properties of NS for $\alpha \approx 1$, we consider Eq. (\ref{17}) to calculate $\Omega_{*}$ rather than Eq. (\ref{23}). Thus, by eliminating $\rho_{c}e^{2}$ using Eqs. (\ref{17}) and (\ref{32}), $Q_{*}$ is given by:
\begin{equation}\label{33}
Q_{*}=-{3 \over 8G}\sqrt{ {2 \over G_{\alpha}} } \left[ 1+{5/3 \over 6-\alpha} \left( {R_{*} \over R_{s}}\right)^{\alpha-1} \right] R_{s}^{5} \Omega_{*}^{2},
\end{equation}
and, in turn, the Love number ($\lambda_{*,\: rot}=\lambda_{*,\: tid}$) reads
\begin{equation}\label{34}
\lambda_{*}={3 \over 8G}\sqrt{ {2 \over G_{\alpha}} } \left[ 1+{5/3 \over 6-\alpha} \left( {R_{*} \over R_{s}}\right)^{\alpha-1} \right] R_{s}^{5}.
\end{equation}
Similar to the $n=0$ polytrope, the dimensionless dynamic properties of NS for the $n=1$ polytropic EoS are as follows:
\begin{equation}\label{35}
\underline{I}_{*}= {2 \over 5}{1 \over C^{2}} \left[ {1+{5/3 \over 6-\alpha} \left( {R_{*} \over R_{s}}\right)^{\alpha-1} \over 1+ {1 \over 4-\alpha} \left( {R_{*} \over R_{s}}\right)^{\alpha-1}} \right],
\end{equation}
for the inertial moment,
\begin{equation}\label{36}
\underline{\lambda}_{*}= {3 \over 8}\sqrt{ {2 \over G_{\alpha}} } {1 \over C^{5}} \left[ 1+{5/3 \over 6-\alpha} \left( {R_{*} \over R_{s}}\right)^{\alpha-1} \right],
\end{equation}
for the Love number, and 
\begin{equation}\label{37}
\underline{Q}_{*}= {75 \over 32}\sqrt{ {2 \over G_{\alpha}} } \left( {1 \over C} \right) { \left[ 1+ {1 \over 4-\alpha} \left( {R_{*} \over R_{s}}\right)^{\alpha-1}\right]^{2} \over \left[ 1+{5/3 \over 6-\alpha} \left( {R_{*} \over R_{s}}\right)^{\alpha-1} \right] },
\end{equation}
for the quadrupole moment. Therefore, the UG I-Love-Q relations for a slowing rotation NS with $n=1$ polytropic EoS are the following:
\begin{equation}\label{38}
\overline{C}_{*,I\lambda}= 4\left( {G_{\alpha} \over 9}\right)^{0.2} {\left[ 1+{5/3 \over 6-\alpha} \left( {R_{*} \over R_{s}}\right)^{\alpha-1} \right]^{0.6} \over \left[ 1+ {1 \over 4-\alpha} \left( {R_{*} \over R_{s}}\right)^{\alpha-1}\right] },
\end{equation}
for the inertial moment and Love number,
\begin{equation}\label{39}
\overline{C}_{*,IQ}= {1024 G_{\alpha} \over 28125} {\left[ 1+{5/3 \over 6-\alpha} \left( {R_{*} \over R_{s}}\right)^{\alpha-1} \right]^{3} \over \left[ 1+ {1 \over 4-\alpha} \left( {R_{*} \over R_{s}}\right)^{\alpha-1}\right]^{5} },
\end{equation}
for the inertial and quadrupole moments, and 
\begin{equation}\label{40}
\overline{C}_{*,Q\lambda} = {75 \over 16} \left( {1 \over 3G_{\alpha}^{2}}\right)^{0.2} { \left[ 1+ {1 \over 4-\alpha} \left( {R_{*} \over R_{s}}\right)^{\alpha-1}\right]^{2} \over \left[ 1+{5/3 \over 6-\alpha} \left( {R_{*} \over R_{s}}\right)^{\alpha-1} \right]^{0.3} },
\end{equation}
for the quadrupole moment and the Love number. As can be seen, the UG I-Love-Q relations for $n=1$ polytrope are different from the usual ones \cite{14}. By setting $\alpha=1$, we can see that $\overline{C}_{*,I\lambda}$ and $\overline{C}_{*,Q\lambda}$ are increased $330\%$ and $70\%$, while $\overline{C}_{*,IQ}$ decreased about $50\%$. The discrepancies in the $\alpha=1$ case is due to simplifications in the density, Eq.(\ref{30}), which approach the $n=0$ polytrope fluid rather than the $n=1$ one. It should be pointed out, in the $n=1$ case, that the way to calculate the dynamical properties of NS differs from the ones of Ref. \cite{14}. 
We shall discuss these issues in Sec. IV.

\section{Results and Discussion}

In this work we apply the UG model for NSs for the ultralow, moderate, and ultrahigh mass regimes, i.e., for the following NSs: 4U1746-37 \cite{38}, M13 \cite{39}, and J0348+0432 \cite{40}, respectively. Table I shows the values of mass ($M_{s}$) and radius ($R_{s}$) in terms of the solar values ($M_{sun}\simeq 1.99\times 10^{30} kg$ and $R_{sun}\simeq 6.96\times 10^{5} km$) together with the relevant observational ranges. 

\vspace{.7 cm}
\begin{center}
{\footnotesize Table I Relevant values for the selected NSs, i.e., 4U1746-37 \cite{38}, M13 \cite{39}, and J0348+0432 \cite{40}.}
\\
\vspace{.5 cm}
\begin{tabular}{|c|c|c|c|c|}
\hline
NS & $(M_{s}\pm\triangle M_{s})/M_{sun}$ & $(R_{s}\pm\triangle R_{s})/R_{sun} \times 10^{-5}$ \\
\hline
4U1746-37 & 0.41 $\pm$ 0.14 & 1.25$\pm$ 0.22 \\
M13 & 1.36$\pm$ 0.04 & 1.42 $\pm$ 0.01 \\
J0438-0432 & 2.01$\pm$ 0.04 & 1.87 $\pm$ 0.29 \\
\hline
\end{tabular}
\end{center}
\vspace{.7 cm}

Based on the uncertainties of the relevant quantities, we obtain bounds on the characteristic length, $R_{*}$, and scaling dimension, $\alpha$, of the UG-inspired model. Firstly, we solve the NHE equation, Eq. (\ref{5}), by the fourth-order Runge-Kutta method for the selected NS's mass and radius. Then, by keeping the relevant polytropic index and core density values, varying $\alpha$ and $R_{*}$ within the UGHE equation, Eq. (\ref{6}), and calculating the same observable parameters, we admit only those values that are compatible with the uncertainties (Table I). We set the polytropic index of $n=0.68$ and the core density $\rho_{c}=6.61\times 10^{14}, 15.06\times10^{14}$, and $9.79\times10^{14} g/cm^{3}$ for 4U1746-37, M13, and J0348+0432, respectively. Figure ~\ref{Fig:UMJ,a-R} depicts the allowed regions of $R_{*}$ and $\alpha$ for the selected NSs. The horizontal and vertical dashed lines are depicted to separate the different regions.
We find that ultralow mass NS admit a wider allowed region in comparison with medium and ultramassive NSs. The ultralow and ultrahigh mass NSs have allowed values of $\alpha$ bigger than the medium ones. It should be emphasized that the obtained bounds on the UG parameters are based on nonrelativistic calculations, Eq. (\ref{6}). 
 
\begin{figure}
\centering
\hfill  
\subfigure [n=0.68] {\includegraphics[width=.4\linewidth]{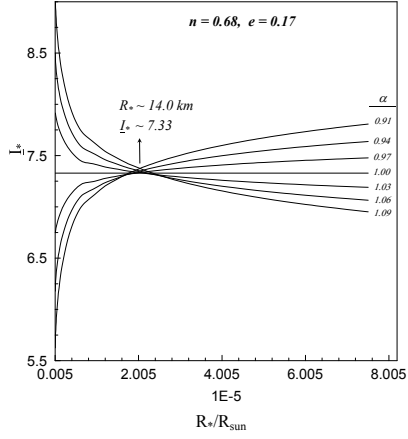}}
\hfill
\subfigure [n=1.00] {\includegraphics[width=.4\linewidth]{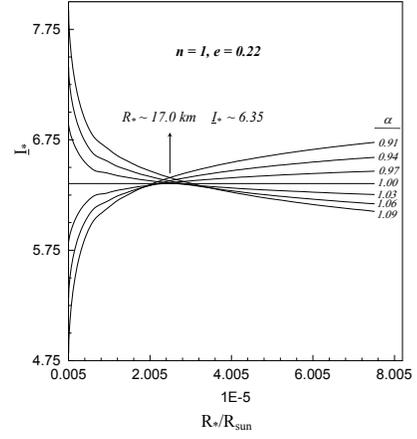}}
\hfill

\caption{The dimensionless UG inertial moment of NS for different allowed values of the characteristic length and scaling dimension. The $R_{*}$ and $\alpha$ values have been limited by the relevant observational data of $M13$ \cite{39} for polytropic indices (a) $n=0.68$ and (b) $n=1.00$.}\label{Fig:IbM13}

\end{figure}

\begin{figure}
\centering
\hfill  
\subfigure [n=0.68] {\includegraphics[width=.4\linewidth]{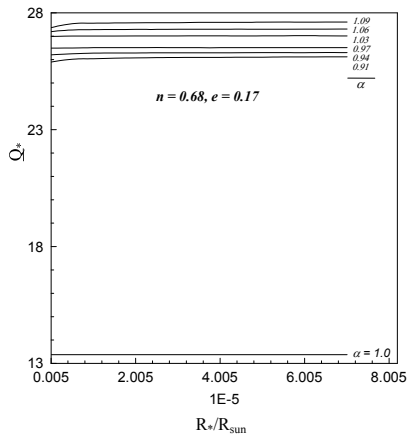}}
\hfill
\subfigure [n=1.00] {\includegraphics[width=.4\linewidth]{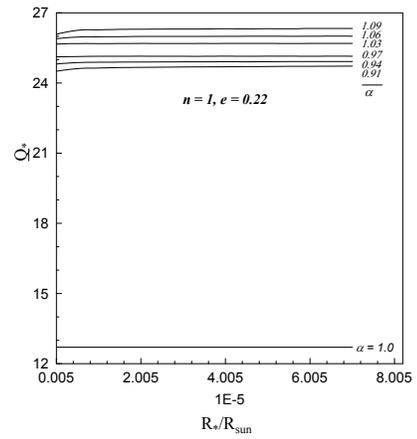}}
\hfill

\caption{The same as Fig. 2 for the UG quadrupole moment.}\label{Fig:QbM13}

\end{figure}

\begin{figure}
\centering
\hfill  
\subfigure [n=0.68] {\includegraphics[width=.4\linewidth]{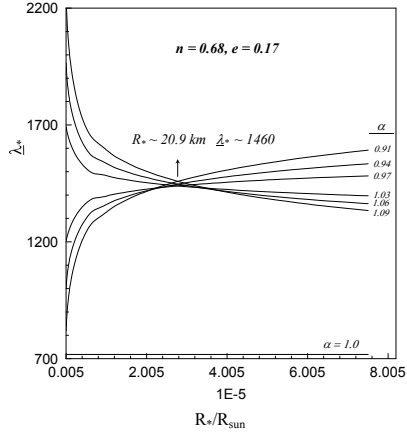}}
\hfill
\subfigure [n=1.00] {\includegraphics[width=.4\linewidth]{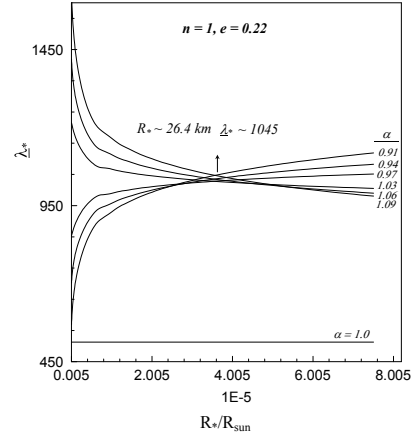}}
\hfill
\caption{The same as Fig. 2 for the UG Love number.}\label{Fig:LbM13}

\end{figure}

In order to examine the I-Love-Q relations in the UG model, we consider the M13 data and calculate $I_{*}$, $\lambda_{*}$, and $Q_{*}$. The assumption that tidal and rotational deformations are exactly the same in the Newtonian limit \cite{14, 48} is kept when considering UG ($\lambda_{*,\: rot}=\lambda_{*,\: tid}$), we do not need to construct the Clairaut-Radau equations to compute tidal deformations. Instead, we obtain rotational deformations by calculating the relevant integral, Eq. (\ref{9}), for which the UG density profile of NS is obtained by solving Eq. (\ref{6}). It is worth noting that $I_{*}$ and $Q_{*}$ are still calculated without any more assumptions at the Newtonian limit. 
In order to get the Love number, we need to calculate $\Omega_{*}$, Eqs. (\ref{17}) and (\ref{23}) for the uniform and nonuniform density cases, respectively. In the $n=0$ case, $I_{*}$ is independent of the UG parameters. However, $Q_{*}$ and $\lambda_{*}$ depend on the $\alpha$ and $R_{*}$ values through $G_{0\alpha}$ since the second term in the bracket of $Maclaurin$ type ungravity potential, Eq. (\ref{13}), is diminished at the surface of the NS when calculating $\Omega_{*}$. In order to match these quantities in the UG model to the universal ones \cite{14}, we set $f_{0}=0$ in Eq. (\ref{15}). Unlike the uniform case, we need to consider the eccentricity of the NS to compute its $Q_{*}$ and $\lambda_{*}$ for nonuniform density. Setting $e=0.17$ and $e=0.22$ for $n=0.68$ and $n=1$, respectively, we obtain the universal values of $\overline{C}_{*,I\lambda}$, $\overline{C}_{*,IQ}$, and $\overline{C}_{*,Q\lambda}$ for the $\alpha=1$ case. We keep these values of eccentricities in the rest of this work. Figures. ~\ref{Fig:IbM13},~\ref{Fig:QbM13}, and ~\ref{Fig:LbM13} show $\underline{I}_{*}$, $\underline{Q}_{*}$, and $\underline{\lambda}_{*}$ of $M13$, respectively, for different values of the allowed values for $\alpha$ and $R_{*}$ for the polytrope $n=0.68$ and $n=1$ EoS. In fact according to Fig. ~\ref{Fig:IbM13}, deviating $\alpha$ from unity by $\sim+/-10\%$ and decreasing $R_{*}$ to the allowed minimum value implies a deviation of $\underline{I}_{*}$ of $25\%$ for $n=0.68$ and $n=1$. We also see that these deviations are symmetric with respect to $\alpha=1$. Figure ~\ref{Fig:QbM13} shows that increasing $R_{*}$  for various $\alpha$ values, renders no significant variation in $\underline{Q}_{*}$. However, we can see that, when deviating $\alpha$ from unity ($\sim+/-10\%$), $\underline{Q}_{*}$ varies significantly ($90\%$ to $110\%$) for both polytrope indices. We can observe that the deviation of $\underline{\lambda}_{*}$ reaches more than $200\%$ when the $\alpha$ increases $\sim10\%$ from unity for both polytropic EoSs at the lowest allowed values of $R_{*}$. Although changes of $\underline{Q}_{*}$ are approximately independent of variations of the characteristic length of the UG model for various scaling dimensions, only for a specific value of $R_{*}$ we get the same value for $\underline{I}_{*}$ and $\underline{\lambda}_{*}$ for different values of $\alpha$.

\begin{figure}
\centering
\hfill  
\subfigure [n=0.68] {\includegraphics[width=.4\linewidth]{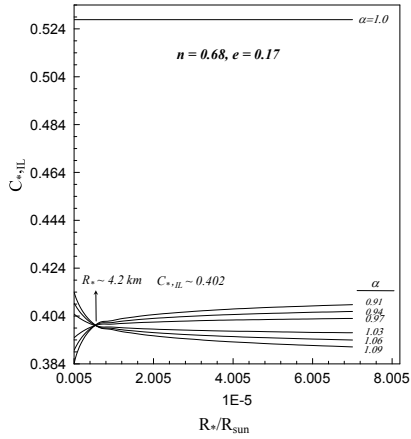}}
\hfill
\subfigure [n=1.00] {\includegraphics[width=.4\linewidth]{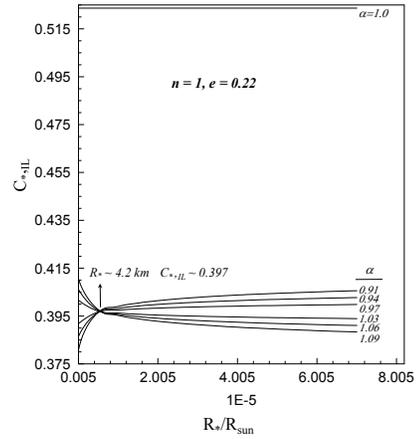}}
\hfill
\caption{The UG I-Love relation for different allowed values of the characteristic length and scaling dimension. The $R_{*}$ and $\alpha$ values have been limited by the relevant observational data of $M13$ \cite{39} for polytropic indices (a) $n=0.68$ and (b) $n=1.00$.}\label{Fig:ILM13}

\end{figure}

\begin{figure}
\centering
\hfill  
\subfigure [n=0.68] {\includegraphics[width=.4\linewidth]{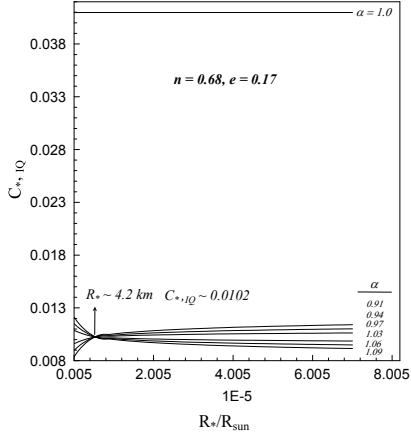}}
\hfill
\subfigure [n=1.00] {\includegraphics[width=.4\linewidth]{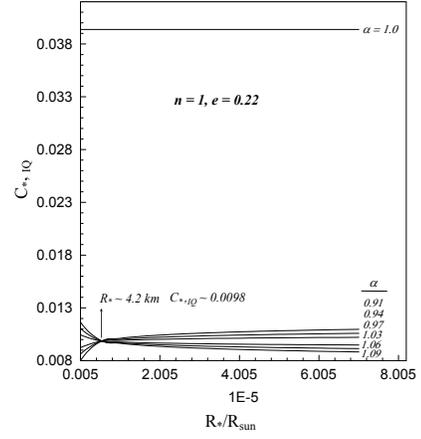}}
\hfill
\caption{The same as Fig. 5 for the UG IQ relation.}\label{Fig:IQM13}

\end{figure}

\begin{figure}
\centering
\hfill  
\subfigure [n=0.68] {\includegraphics[width=.4\linewidth]{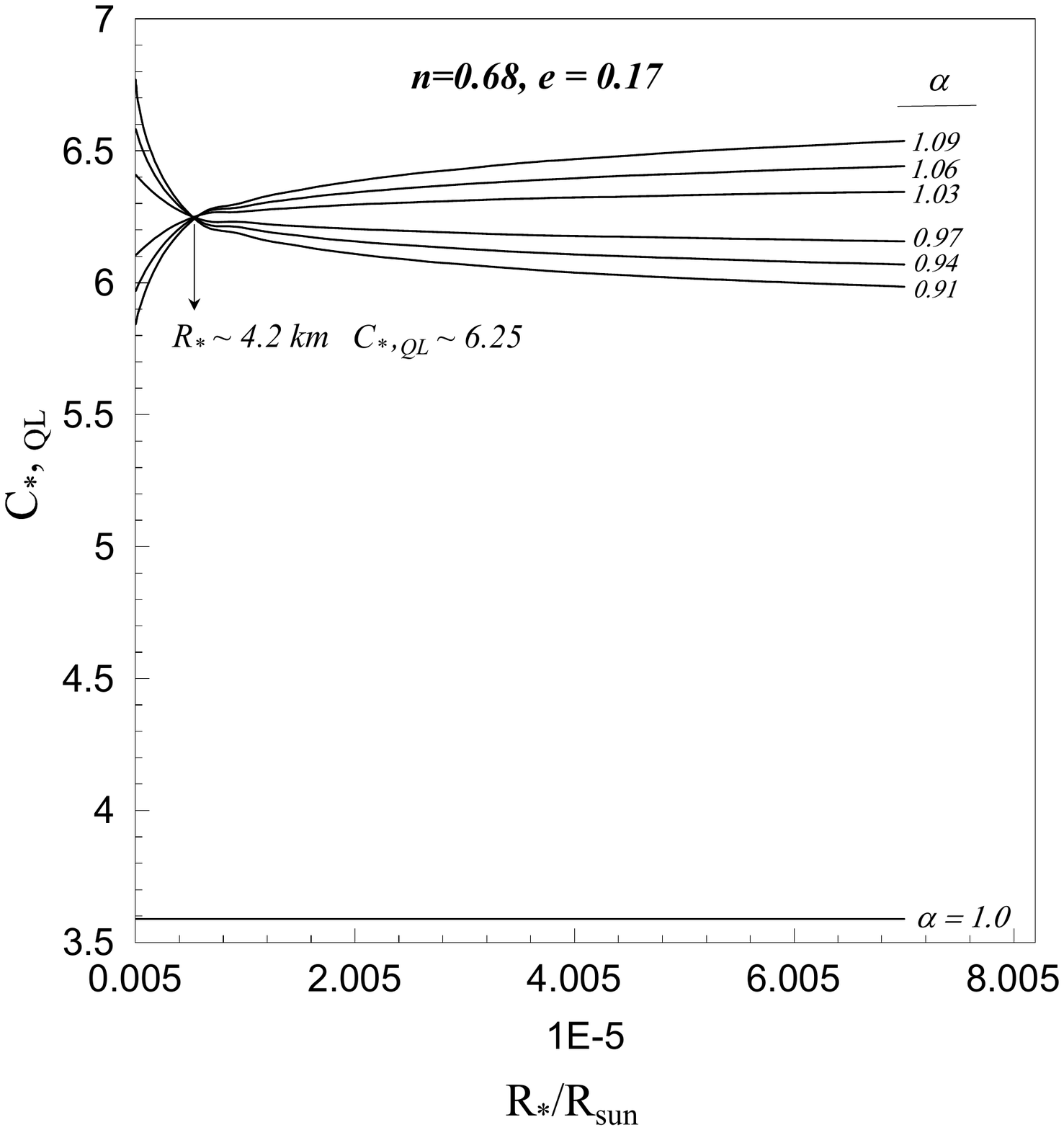}}
\hfill
\subfigure [n=1.00] {\includegraphics[width=.4\linewidth]{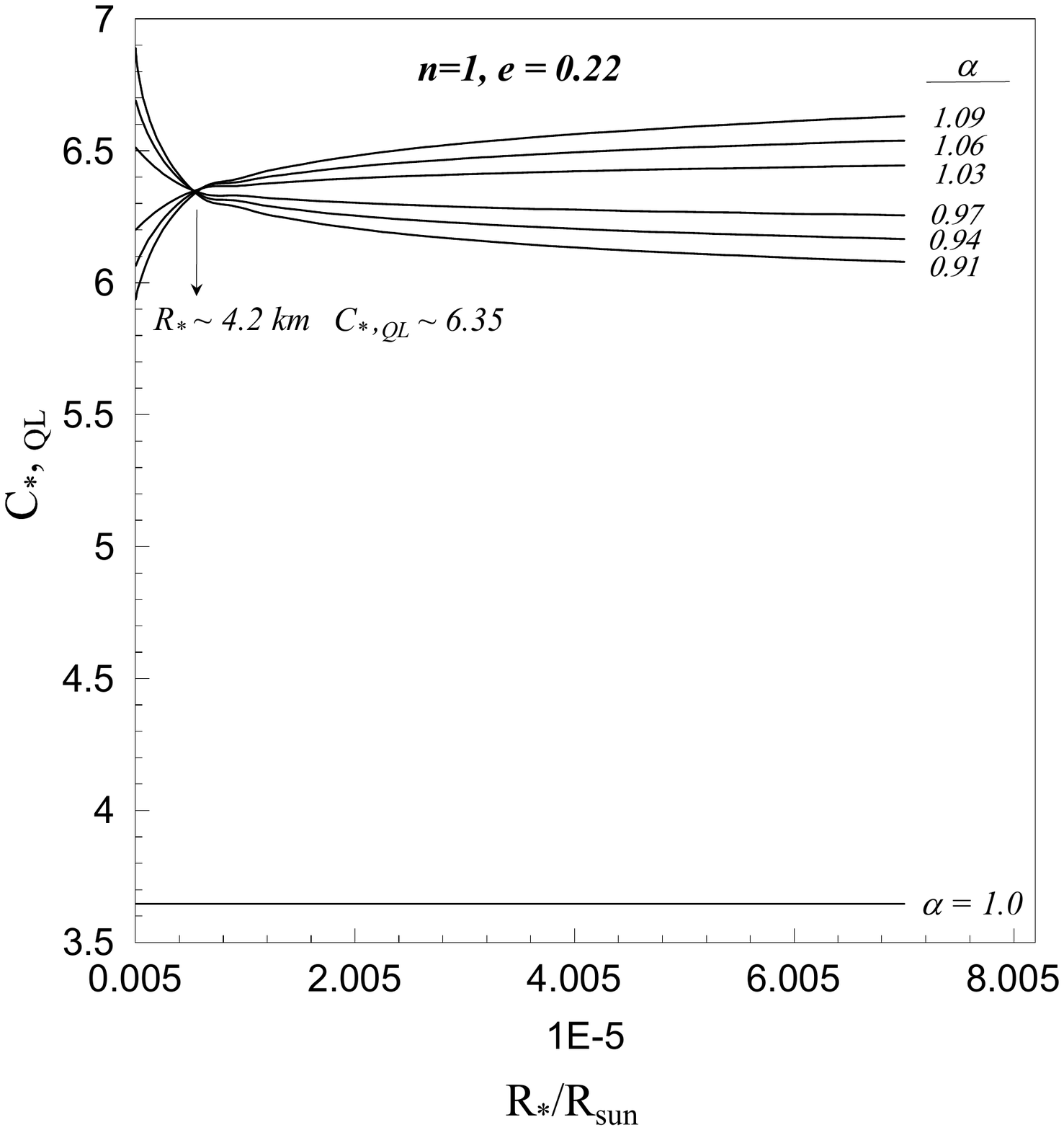}}
\hfill

\caption{The same as Fig. 5 for the UG Q-Love relation.}\label{Fig:QLM13}
\end{figure}

Finally, we calculate $ \overline{C}_{*,I \lambda} $, $ \overline{C}_{*,IQ} $, and $ \overline{C}_{*,Q \lambda} $, by using Eq. (\ref{12}) for the different allowed values of $ \alpha $ and $ R_{*} $. Figures ~\ref{Fig:ILM13}, ~\ref{Fig:IQM13} and ~\ref{Fig:QLM13} depict the UG I-Love-Q relations versus $R_{*}$ for different $ \alpha $ and for polytropic indices $ n=0.68 $ and $ n=1 $. Although there is no significant dispersion in the UG I-Love-Q relations for various $\alpha$'s and for different allowed values of $ R_{*} $, there are deviations of about $ 30\% $, $ 300\% $, and $ 70\% $ for $\overline{C}_{*,I \lambda}$, $\overline{C}_{*,IQ}$, and $\overline{C}_{*,Q\lambda}$, respectively, from the universal values \cite{14}. We find the UG I-Love-Q relations have fixed values independent of the scaling dimension value at $ R_{*}\sim 4.2 \ km$. All relevant figures of the dynamic properties of $M13$ show that there are significant gaps between $dimensionless$ dynamic quantities when UG is switched on except in what concerns the dimensionless inertial moment. It seems that the discrepancies are originated from $\Omega_{*}$ which influences $\underline{Q}_{*}$ through $\chi_{*}$, a dimensionless measure of the NS's angular momentum [cf. Eq. (\ref{11})]. Our numerical calculations show that when the value of $\alpha$ changes $ \pm 0.5\% $, the absolute values of $Q_{*}$ increase only $ 2\% $, while the angular velocity values decrease about $ 30\% $ and, in turn, the values of $\underline{Q}_{*}$ and $\underline{\lambda}_{*}$ increase about $ 100\% $ and $ 90\% $, respectively.

\vspace{.7 cm}
\noindent

In conclusion, we have investigated the UG hydrostatic equilibrium equations in the framework of polytropic EoS for the selected pulsars at the ultralow, moderate, and ultrahigh mass regimes and gotten bounds on the UG parameters, i.e., the scaling dimension and the characteristic length, based on their observational ranges of masses and radii. We have analytically and numerically examined the universal I-Love-Q relations at the Newtonian limit for the uniform and nonuniform NSs in the framework of the UG model. In order to get the universal I-Love-Q relations for the uniform NS, the gravity constant should be kept in the presence of UG. In the nonuniform case, switching on UG instantly leads to a significant change in the quadrupole moment and Love number of NS rather than its inertial moment. We have also found that the UG I-Love-Q relations exhibit a clear deviation from the usual ones.
\vskip 0.5cm

\noindent
{\bf ACKNOWLEDGMENTS}

\vskip 0.5cm

\noindent

H.M. wishes to thank the organizers and speakers of NewCompStar School 2016, a COST initiative, September 5–9, 2016, Department of Physics of the University of Coimbra, Portugal.

\appendix

\section{The value of $\Omega_{*}$ in the case of an $n=1$ polytrope}
In the case of an $n=1$ polytrope of a slowly rotating fluid in the presence of the UG potential, $\phi_{*}(r)$, let us consider a spheroid which rotates around the $z$ axis with a constant angular velocity, $\Omega_{*}$.

We introduce the centrifugal potential, $\phi_{*c}$, in such a way that we have $\nabla \phi_{*c}={d\vec v_{*} \over dt}$ (keeping the Roche model \cite{48} in the case of UG). Since $\vec v_{*}=\vec{\Omega}_{*}\times \vec{r}$, we have ${d\vec v_{*} \over dt}=-\Omega_{*}^{2} (\vec{x}+\vec{y})$. Then we obtain
\begin{equation}\label{B1}
\phi_{*c}=-{1 \over 2}\Omega_{*}^{2}(x^{2}+y^{2}).
\end{equation}
At the UGHE situation, we have
\begin{equation}\label{B2}
{d\vec v_{*} \over dt}=-{1 \over \rho_{*}(r)} \nabla P_{*}(r) - \nabla \Psi_{*}(r)=0, \quad  \Psi_{*}(r)=\phi_{*}(r)+\phi_{*c}(r). \qquad [Eq. \: (\ref{22})]
\end{equation}
Recalling that $P_{*}=K\rho_{*}^{1/n+1}$ with $n=1$ and doing the integral over the above equation, we obtain
\begin{equation}\label{B3}
2{P_{*}(r) \over \rho_{*}(r)} + \Psi_{*}(r)=C_{0*},
\end{equation}
where $C_{0*}$ is a constant whose value we set the same as that in the nonrotating case ($\Omega_{*}=0$). Since $P_{*}(R_{s})=0$, we have $C_{0*}=\phi_{*}(R_{s})$ or 
\begin{equation}\label{B4}
C_{0*}=-{G_{\alpha}M \over R_{s}} \left[ 1+\left( {R_{*} \over R_{s}}\right) ^{\alpha-1}\right] .
\end{equation}
 Since our goal is finding the maximum value of $\Omega_{*}$ at the equator of the NS, we can write
\begin{equation}\label{B5}
[{d\Psi_{*} \over dr}]_{R_{*c}}=0.
\end{equation}
Doing the derivative $\phi_{*}(r)$ and $\phi_{*c}$, we obtain 
\begin{equation}\label{B6}
\Omega_{*c}^{2}={G_{\alpha} M \over R_{*c}^{3}}\left[1+\alpha\left({R_{*} \over R_{*c}}\right)^{\alpha-1} \right],
\end{equation}
where we should still obtain $R_{*c}$. According to Eq. (\ref{B3}), the value of $\Psi_{*}(R_{*c})$ will be maximum when $\left[{P_{*} \over \rho_{*}}\right]_{R_{*c}}=0$. Thus, similar to the gravity case \cite{48}, we reach the equality $\Psi_{*}(R_{*c})=C_{0*}$ and after a bit of manipulation we obtain
\begin{equation}\label{B7}
\left[R_{*c}-{3 \over 2}R_{s} \right]+\left[ \left( {R_{*} \over R_{s}}\right)^{\alpha-1}R_{*c}-\left( {R_{*} \over R_{*c}}\right)^{\alpha-1}\left( {\alpha+2 \over 2 }\right) R_{s} \right]=0.  
\end{equation}
If the first bracket equals zero we obtain a trivial solution like the usual gravity case, that is, $R_{*c}={3 \over 2}R_{s}$ \cite{48}. Setting the second bracket to zero leads to 
\begin{equation}\label{B8}
R_{*c}=\left( {\alpha+2 \over 2}\right)^{{1 \over \alpha}} R_{s},
\end{equation}
where if we set $\alpha=1$, we again obtain the solution of the usual gravity case. Now, including Eq. (\ref{B8}) into Eq. (\ref{B6}) and using straightforward mathematics, we obtain 
\begin{equation}\label{B9}
\Omega_{*}^{2}=G_{\alpha}\left[1+\alpha\left({2\over 2+\alpha}\right)^{{\alpha-1\over \alpha}}\left({R_{*}\over R_{s}}\right)^{\alpha-1}\right] \left({2\over2 + \alpha}\right)^{{3\over\alpha}} \left( {GM_{s} \over R_{s}^{3}}\right).  
\end{equation}

\pagebreak

\begin{figure}
\centering
\hfill  
\subfigure [4U1746-37] {\includegraphics[width=.4\linewidth]{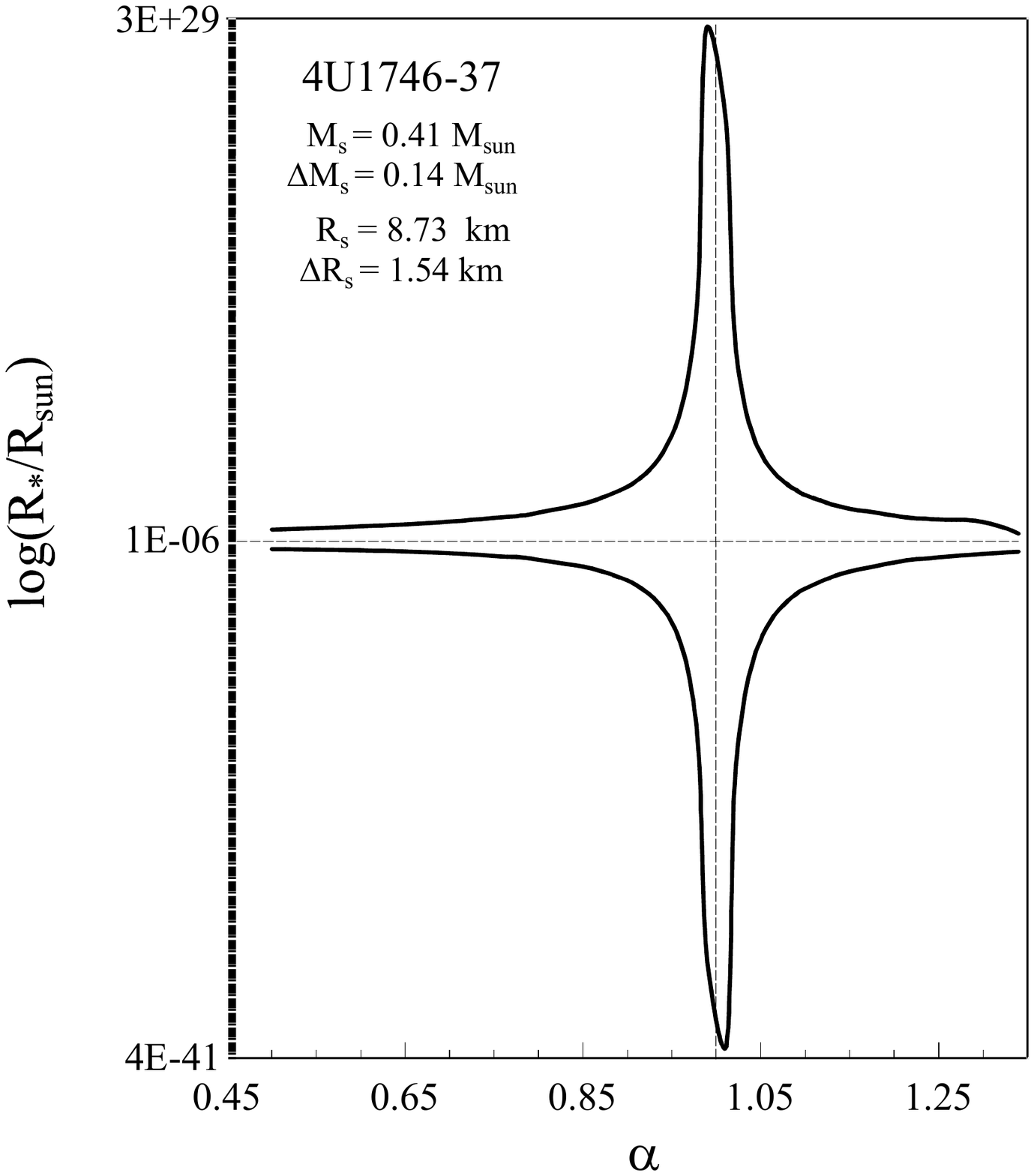}}
\hfill
\subfigure [M13] {\includegraphics[width=.4\linewidth]{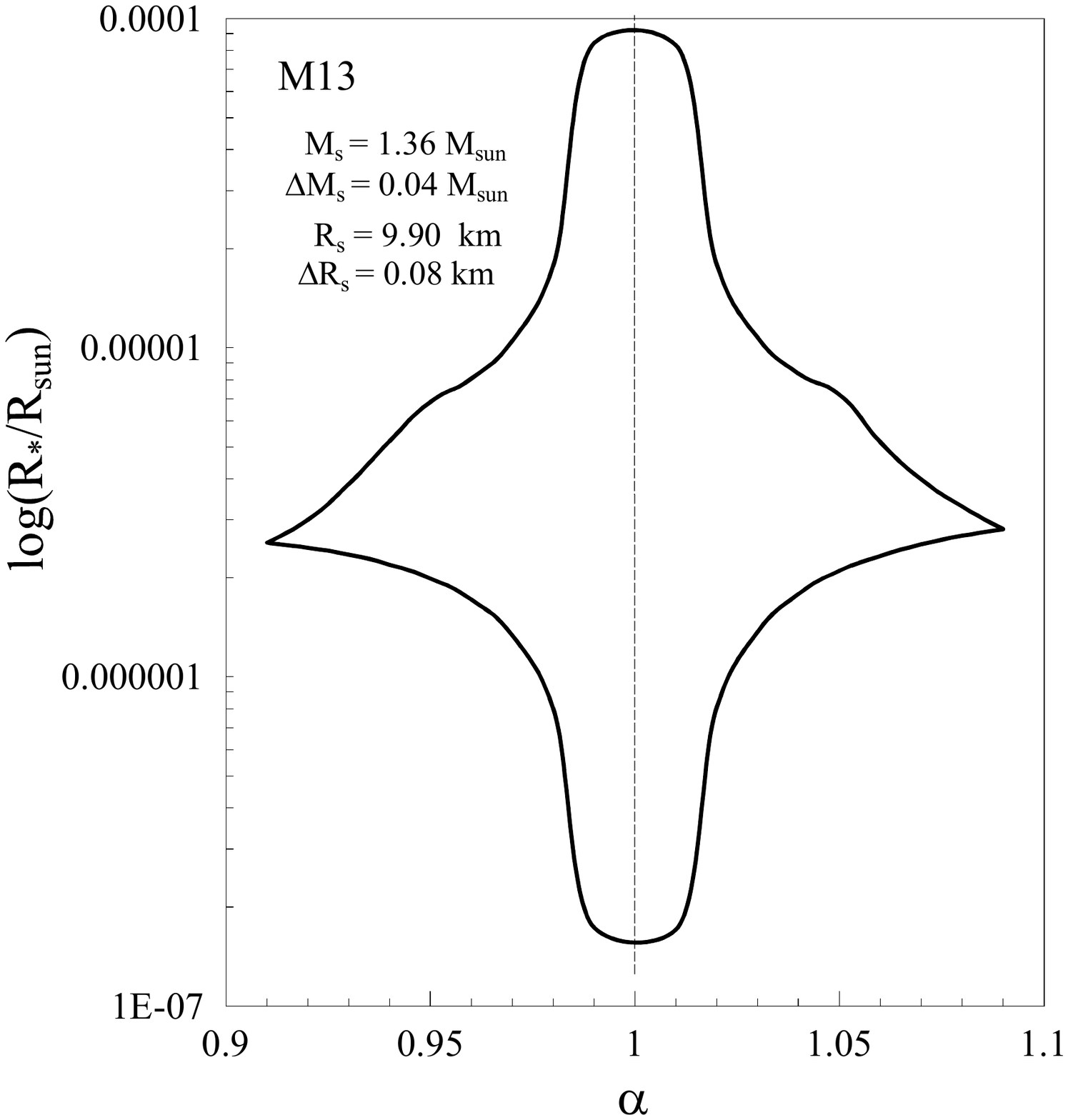}}
\hfill
\subfigure [J0438+0432] {\includegraphics[width=.4\linewidth]{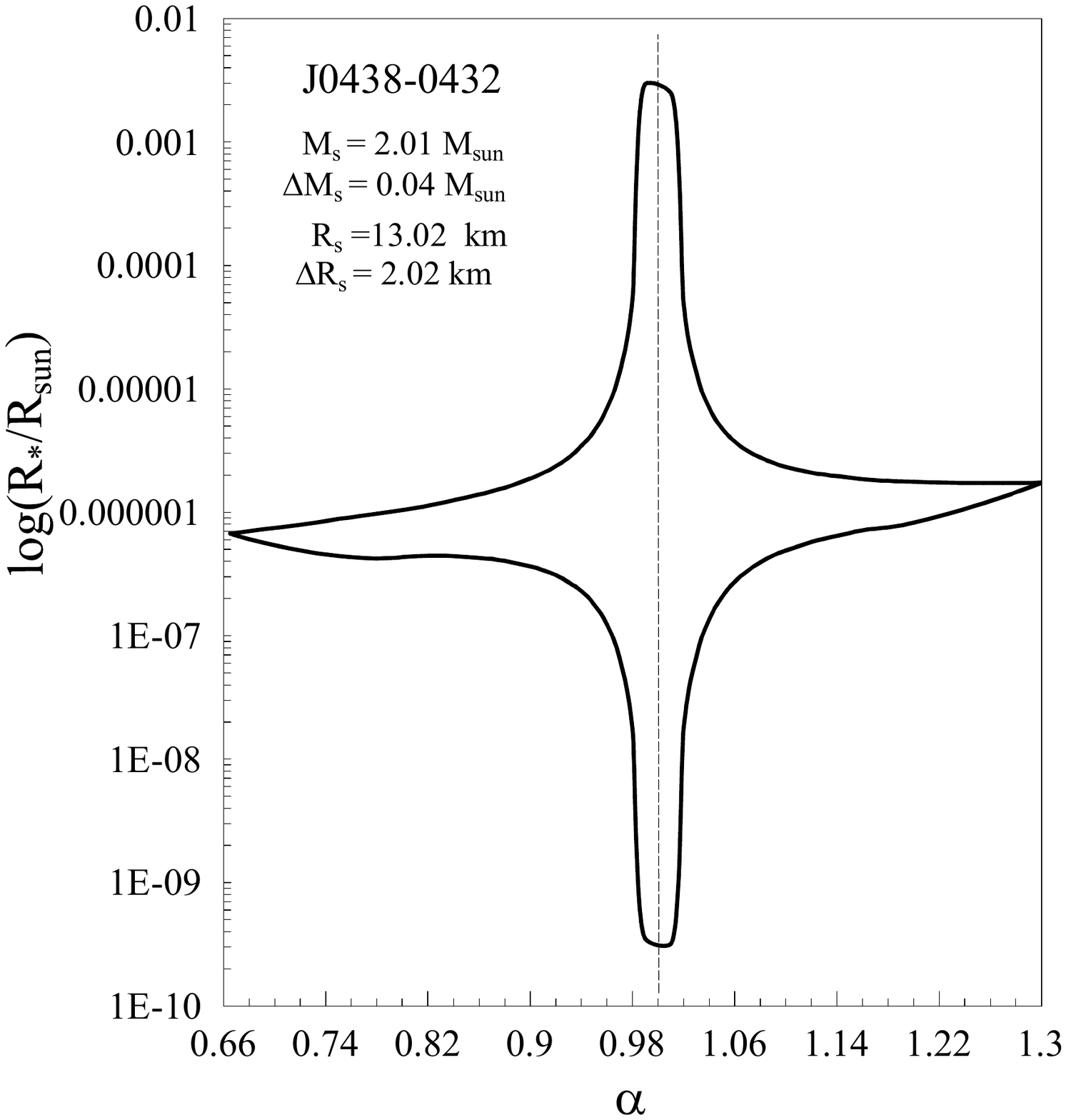}}
\hfill
\hfill
\caption{The allowed region for the UG parameters $\alpha$ and $R_{*}$ for (a) 4U1746-37 \cite{38}, (b) M13 \cite{39}, and (c) J0348+0432 \cite{40} with the polytropic index $n=0.68$.}\label{Fig:UMJ,a-R}
\end{figure}

\begin{thebibliography}{99}

\bibitem{1} D. Lonardoni, A. Lovato, S. Gandolfi, and F. Pederiva, Phys. Rev. Lett. {\bf 114}, 092301 (2015).     
\bibitem{2} C.J. Horowitz, D.K. Berry, C.M. Briggs, M.E. Caplan, A. Cumming, and A.S. Schneider, Phys. Rev. Lett. {\bf 114}, 031102 (2015).
\bibitem{3} S. Kisaka, K. Ioka, and T. Nakamura, Astrophys.J. {\bf 809}, 1 (2015).
\bibitem{4} E. Berti, \textit{et al.}, Class. Quant. Grav. {\bf 32}, 243001 (2015).
\bibitem{5} B. Margalit, B.D. Metzger, and A.M. Beloborodov, Phys. Rev. Lett. {\bf 115}, 171101(2015).
\bibitem{6} S. Bernuzzi, T. Dietrich, and A. Nagar, Phys. Rev. Lett. {\bf 115}, 091101 (2015).
\bibitem{7} K. Takami, L. Rezzolla, and L. Baiotti, Phys. Rev. Lett. {\bf 113}, 091104 (2014).
\bibitem{8} Y.-H. Sham, L.-M. Lin, and P.T. Leung, Astrophys. J. {\bf 781}, 66 (2014).
\bibitem{9} K. Yagi and N. Yunes, Phys. Rep. {\bf 681}, 1 (2017).
\bibitem{10} K. Yagi and N. Yunes, Science {\bf 341} 365 (2013).
\bibitem{11} J. Aasi \textit{et al.} (LIGO Scientific Collaboration), Class. Quantum Grav. {\bf 32}, 074001(2015).
\bibitem{12} F. Acernese \textit{et al.} (VIRGO Collaboration), Class. Quantum Grav. {\bf 32}, 024001 (2015).
\bibitem{13} Y. Aso, Y. Michimura, K. Somiya, M. Ando, O. Miyakawa,T. Sekiguchi, D. Tatsumi, and H. Yamamoto, Phys. Rev. D {\bf 88}, 043007 (2013).
\bibitem{14} K. Yagi, N. Yunes, Phys. Rev. D {\bf 88}, 023009 (2013).
\bibitem{15} T. Hinderer, \textit{et al.}, Phys. Rev. Lett. {\bf 116}, 181101 (2016).
\bibitem{16} K. Kawaguchi, K. Kyutoku, H. Nakano, H. Okawa, M. Shibata, and K. Taniguchi, Phys. Rev. D {\bf 92}, 024014 (2015).
\bibitem{17} H. Goldberg and P. Nath, Phys. Rev. Lett. {\bf 100}, 031803 (2008).
\bibitem{18} H. Georgi, Phys. Rev. Lett. {\bf 98}, 221601 (2007).
\bibitem{19} H. Georgi, Phys. Lett. B {\bf 650}, 275 (2007).
\bibitem{20} K. Cheung, W.-Y. Keung, and T.-C. Yuan, Phys. Rev. Lett. {\bf 99}, 051803 (2007).
\bibitem{21} K. Cheung, W.-Y. Keung, and T.-C. Yuan, Phys. Rev. D. {\bf 76}, 055003 (2007).
\bibitem{22} Y. Liao and J. Y. Liu, Phys. Rev. Lett. {\bf 99}, 191804 (2007).
\bibitem{23} M. Luo and G. Zhu, Phys. Lett. B {\bf 659}, 341 (2008).
\bibitem{24} A. Freitas and D. Wyler, Astro unparticle physics, J. High Energy Phys. {\bf 12} (2007) 033.
\bibitem{25} H. Davoudiasl, Phys. Rev. Lett. {\bf 99}, 141301 (2007).
\bibitem{26} P. K. Das, Phys. Rev. D {\bf 76}, 123012 (2007).
\bibitem{27} S. Hannestad, G. Raffelt, and Y. Y. Y. Wong, Phys. Rev. D {\bf 76}, 121701 (2007).
\bibitem{28} G. L. Alberghi, A. Y. Kamenshchik, A. Tronconi, G. P. Vacca, and G. Venturi, Phys. Lett. B {\bf 662}, 66 (2008).
\bibitem{29} S. Das, S. Mohanty, and K. Rao, Phys. Rev. D {\bf 77}, 076001 (2008).
\bibitem{30} T. Kikuchi and N. Okada, Phys. Lett. B {\bf 665}, 186 (2008).
\bibitem{31} J. McDonald, J. Cosmol. Astropart. Phys. {\bf 03} (2009) 019.
\bibitem{32} J. R. Mureika, Phys. Rev. D {\bf 79}, 056003 (2009).
\bibitem{33} O. Bertolami and N. M. C. Santos, Phys. Rev. D {\bf 79}, 127702 (2009).
\bibitem{34} J. Mureika and E. Spallucci, Phys. Lett. B {\bf 693}, 129 (2010).
\bibitem{35} O. Bertolami, J. P\'aramos, and P. Santos, Phys. Rev. D {\bf 80}, 022001 (2009).
\bibitem{36} O. Bertolami and H. Mariji, Phys. Rev. D {\bf 93}, 104046 (2016).
\bibitem{37} R. A. de Souza and J.E. Horvath, Phys. Rev. D {\bf 86}, 027502 (2012).
\bibitem{38} Z. Li, Z. Qu, L. Chen, Y. Guo, J. Qu, and R. Xu, Astrophys. J. {\bf 798}, 56 (2015).
\bibitem{39} A. Catuneanu, C. O. Heinke, G. R. Sivakoff, W. C. G. Ho, and M. Servillat, Astrophys. J. {\bf 764}, 145 (2013). 
\bibitem{40} J. Antoniadis \textit{et al.}, Science {\bf 340}, 1233232 (2013).
\bibitem{41} E. G. Adelberger, B.R. Heckel, S. Hoedl, C.D. Hoyle, D.J. Kapner, and A. Upadhye, Phys. Rev. Lett. {\bf 98}, 131104 (2007).
\bibitem{42} O. Bertolami, F. Francisco, and P.J.S. Gil, Class. Quant. Grav. {\bf 33}, 125021 (2016).
\bibitem{43} N. G. Deshpande, S. D. H. Hsu and J. Jiang, Phys. Lett. B {\bf 659}, 888 (2008).
\bibitem{44} D.-C. Dai, So. Dutta, and D. Stojkovic, Phys. Rev. D {\bf 80}, 063522 (2009).
\bibitem{45} P. Nicolini, Phys. Rev. D {\bf 82}, 044030 (2010).
\bibitem{46} J. R. Oppenheimer and G. M. Volkoff, Phys. Rev. {\bf 55}, 374 (1939).
\bibitem{47} S. Shapiro and S. Teukolsky, {\it Black Hole, White Dwarfs and Neutron Stars: The Physics of Compact Objects} (Wiley-VCH, Weinheim, 1983).
\bibitem{48} T. Mora and C. M. Will, Phys. Rev. D {\bf 69}, 104021 (2004).
\bibitem{49} T. Hinderer, Astrophys. J. {\bf 677}, 1216 (2008).
\bibitem{50} W. G. Laarakkers and E. Poisson, Astrophys. J. {\bf 512}, 282 (1999).

\end{thebibliography}
\end{document}